\documentclass[sigconf]{acmart}
\usepackage{etoolbox}
\patchcmd{\thebibliography}
  {\settowidth}
  {\setlength{\itemsep}{0pt plus 0.1pt}\settowidth}
  {}{}
\apptocmd{\thebibliography}
  {\small}
  {}{}

\usepackage{graphicx}
\usepackage{subfigure}
\usepackage{colortbl}
\usepackage{framed}
\usepackage{comment}
\usepackage{url}
\usepackage{hyperref}
\usepackage{flushend}
\usepackage{balance}

\usepackage{url}
\AtBeginDocument{%
  \providecommand\BibTeX{{%
    \normalfont B\kern-0.5em{\scshape i\kern-0.25em b}\kern-0.8em\TeX}}}

\setcopyright{acmcopyright}
\copyrightyear{2022}
\acmYear{2022}
\acmDOI{XXXXXXX.XXXXXXX}

\acmConference[SEAMS 2022]{The 17th Symposium on Software Engineering for Adaptive and Self-Managing Systems}{May 21–29, 2022}{Pittsburgh, PA, USA}
\acmPrice{15.00}
\acmISBN{978-1-4503-XXXX-X/18/06}

\begin{document}

\title{Preliminary Results of a Survey\\on the Use of Self-Adaptation in Industry}



\author{Danny Weyns$^{1,3}$, 
Ilias Gerostathopoulos$^{2}$, 
Nadeem Abbas$^{3}$, 
Jesper Andersson$^{3}$ 
\\Stefan Biffl$^{4}$, 
Premek Brada$^{5}$, 
Tomas Bures$^{6}$, 
Amleto Di Salle$^{7}$ 
\\Patricia Lago$^{2}$, 
Angelika Musil$^{4,1}$, 
Juergen Musil$^{4}$ 
and Patrizio Pelliccione$^{8}$}

\affiliation{%
  \institution{$^{1}$Katholieke Universiteit Leuven; $^{2}$Vrije Universiteit Amsterdam; $^{3}$Linnaeus University; $^{4}$CDL-SQI, TU Wien; $^{5}$University of West Bohemia, Pilsen; $^{6}$Charles University, Prague; $^{7}$University of L'Aquila; $^{8}$Gran Sasso Science Institute, L'Aquila}
  \country{$^{1}$Belgium, $^{2}$The Netherlands, $^{3}$Sweden, $^{4}$Austria, $^{5}$Czech Republic, $^{7}, ^{8}$Italy}
}

\email{
Contact: danny.weyns@kuleuven.be,  i.g.gerostathopoulos@vu.nl
}\vspace{20pt}
\renewcommand{\shortauthors}{D. Weyns et al.}
\renewcommand{\shorttitle}{Preliminary Results of a Survey on the Use of Self-Adaptation in Industry}
\begin{abstract}
Self-adaptation equips a software system with a feedback loop that automates tasks that otherwise need to be performed by operators. Such feedback loops have found their way to a variety of practical applications, one typical example is an elastic cloud. Yet, the state of the practice in self-adaptation is currently not clear. To get insights into the use of self-adaptation in practice, we are running a large-scale survey with industry. This paper reports preliminary results based on survey data that we obtained from 113 practitioners spread over 16 countries, 62 of them work with concrete self-adaptive systems. We highlight the main insights obtained so far: motivations for self-adaptation, concrete use cases, and difficulties encountered when applying self-adaptation in practice. We conclude the paper with outlining our plans for the remainder of the study.
\end{abstract}



\maketitle

\section{Introduction}\label{sec:introduction}


Software-intensive systems are systems where software plays a vital role in the construction and operation of these systems~\cite{Holzl2008}.
Software-intensive systems form the backbone of our factories, traffic, healthcare, telecommunication, finances, and so forth. The trustworthiness and sustainability of these systems is therefore vital for our society.
Yet, achieving this is challenging due to complexity of these systems. Our focus here is on the challenge of these systems when dealing with uncertain conditions they face during operation.

A classic approach to deal with uncertainty is equipping a software-intensive system with a feedback loop that automates tasks that are otherwise performed by operators. The purpose of a feedback loop can be very diverse, ranging from minimising the cost of operation under workloads that are hard to predict, ensuring a required level of performance under changing network conditions, dealing with errors caused by external services, or defending the system against malicious attacks and the problems they may cause. 
Applying feedback loops to software-intensive systems has been subject of active research in academics. Back in 1998, Oreizy et al.~\cite{oreizy1999aba} introduced the notion of \textit{self-adaptation} that comprises two interacting processes: one that detects and responds to changes, and the other that deals with evolution of the system over time. Garlan et al.~\cite{garlan2004rainbow} pointed out the crucial role of first-class architectural models that enable a system to reason about system-wide change to adapt and satisfy its goals. 
Kramer and Magee~\cite{Kramer2007SMS} highlighted the role of software architecture in defining self-adaptive systems, distinguishing adaptation management from goal management. The last decade, the research community developed a vast body of knowledge and know-how on engineering self-adaptive systems,  e.g., in research road maps~\cite{cheng2009software,Lemos2013roadmap,978-3-319-74183-3_1,Assmann2014}, books~\cite{lalanda2013,weyns2021introduction} and literature studies, e.g.,~\cite{Weyns2013,8787041,Musil2017,Omid2021,Alfonso2021,Moghaddam2020SelfadaptiveMS,wong2021selfadaptive}. Yet, only a few joint efforts in collaboration with industry are reported, e.g.,~\cite{Hurtado2011,6595488,7968147,978-3-030-00761-4,Javier2016,Bolender21,Weyns2022}.

In parallel, the principles of feedback control were studied and applied by industry. Around 2000, IBM launched its legendary initiative on  autonomic computing~\cite{kephart2003vision}. The central idea of autonomic computing was to enable computing systems to manage themselves based on high-level goals, similar to the human body. Four classic goals are self-optimisation, self-healing, self-protection, and self-configuration. Autonomic computing delegates the complex tasks that traditionally need to be performed by operators to the machine. Over the years, solutions based on feedback loops have found their way to practical applications, for instance, in the domains of elastic clouds and automated management of server parks, see e.g.,\,\cite{beyer2016,spyker2020}.

So on the one hand, academics have established a vast body of knowledge on engineering self-adaptive systems. On the other hand, the principles of feedback loops are applied by industry. While the output of academic research is documented, the current use of self-adaptation in industry and state of the practice is not clear. 

To get insight into the use of self-adaptation in industry, we are running a large-scale survey with active practitioners. This survey aims at shining light on what motivates practitioners to apply self-adaptation, what kind of problems they solve using self-adaptation, how they solve these problems, whether they have any established practices, what risks and challenges they face in adopting self-adaptation, and what opportunities industry sees in this area. Such insights will help the research community to align their efforts with industrial needs, and they help  practitioners by clarifying benefits as well as issues with applying self-adaptation and pointing out opportunities for tackling open challenges.
To the best of our knowledge, no systematic study has been done that investigates and answers questions like those specified above. 

In this paper, we report preliminary results of the survey based on responses of 62 practitioners from 16 countries that work with concrete self-adaptive systems. We highlight the main insights obtained so far: motivations for self-adaptation, concrete use cases, and difficulties encountered when applying self-adaptation in practice. 
We target the following three research questions:

\begin{itemize}
    \item[RQ1:] What motivates practitioners to apply self-adaptation?   
    \item[RQ2:] What are concrete use cases of  self-adaptation in practice?
    \item[RQ3:] What difficulties do practitioners face when applying self-adaptation?
\end{itemize}

Investigating these questions will give us initial insights in whether research and practice in self-adaptation are aligned on the studied topics and whether there are differences in emphasis. 

The remainder of this paper is structured as follows. In Section~\ref{sec:research_method}, we summarise the study design and list the survey questions. Section~\ref{sec:results} presents the results, grouped per research question. We discuss threats to validity in  Section~\ref{sec:discussion}. Finally, we draw conclusions and outline our plans for the rest of the study in Section~\ref{sec:conclusions}.

\section{Study Design}
\label{sec:research_method} 

We present the target population and sampling, the survey instrument, the survey questions, and data analysis. For a full replication package with all study material, the raw data, and the analysis results, we refer to the study website~\cite{replication-package}.
%
%
\vspace{5pt}\\\noindent \textbf{Population and Sampling}. Our target population are practitioners actively involved in the engineering of industrial software-intensive systems in any domain. This includes architects, designers, developers, testers, maintainers, operators, among others who have technical expertise in the engineering of software systems. 

Concretely, we contacted 203 practitioners from a wide variety of companies via the networks of the researchers involved in this study to complete the survey. We used two criteria to invite people: (1) a good representation of domains of the current landscape of software-intensive systems, and (2) participants have the required expertise to answer the questions. The invited practitioners were spread over in total 16 countries.\footnote{Czech Republic 40, Sweden 35, Austria 29, Belgium 28, Germany 18, The Netherlands 16, Spain 9, Denmark 7, UK and USA each 5, France 3, Switzerland 2, and Greece, Poland, Norway, and Japan each 1} The invitations were sent by personalised emails on November 30, 2020. We sent reminders according to a predefined schedule after one, two, and six weeks. 
%
%
\vspace{5pt}\\\noindent \textbf{Survey Instrument}. A survey collects data based on a set of predefined questions~\cite{gray2013drr}. We used closed questions that have a predefined set of possible answers that participants can pick from, and open questions that provide space that respondents can use to explain an answer in detail. While closed questions allow getting a clear view on a particular topic using basic statistics, open questions allow getting detailed in-depth insights using qualitative analysis. 

We used a self-administered anonymous online questionnaire (Survey \& Report hosted by Linnaeus University), allowing to involve a large set of respondents with relatively low cost. 
We created an initial list of survey questions that were directly derived from the research questions of the study. We anticipated that completing the questionnaire would take approximately half an hour. 

We validated the questionnaire in a pilot with eight randomly selected participants from the target population. For this pilot, we added additional meta-questions to the questionnaire about clarity of terminology, relevance and scope of the questions, and the time required to complete the survey. For both clarity of terminology and clarity of the questions we obtained an average score of 4.38 on a scale from 1 (Not clear at all) to 5 (Very clear). None of the participants indicated that questions should be removed or modified. The average reported time to complete the survey was 24 minutes. Based on the feedback, we made a few adjustments in the introductory part of the questionnaire and moved one of the questions up in the list. The finalised questionnaire was then distributed to the participants as explained above. 
%
%
\vspace{5pt}\\\noindent \textbf{Survey Questions}.
We summarise here the questions that are relevant to the preliminary results reported in this paper. For the complete questionnaire we refer to the study website~\cite{replication-package}.

The first set of questions probed whether the participant has worked with self-adaptive systems (Q0.1) and solicited demographic information about the domain (Q0.2), the size of the company (Q0.3), the participant's role (Q0.4) and expertise (Q0.5).  
The second set provides questions related to RQ1 collecting data about the types of problems for which the respondents apply self-adaptation (Q1.1), and the benefits they obtained from applying self-adaptation (Q1.2). 
The third set had a single question related to RQ2 on use cases for self-adaptation. The question asked the respondents to describe a concrete self-adaptive system they worked with (Q2.1).  
Finally, the fourth set addressed RQ3 on difficulties and challenges with applying self-adaptation. The first question investigates the frequency of difficulties (Q3.1); the second question elicited concrete examples of difficulties experienced by the participant (Q3.2). The concrete questions are further explained in the results section. 
%
%
\vspace{5pt}\\\noindent \textbf{Data Analysis}. 
To analyse closed questions, we used quantitative data analysis. We report percentages relative to the respective number of responses, and frequency analyses. 
To analyse textual answers, we used coding~\cite{Strauss1990}. We developed codes and inferred categories from the data by labelling small coherent fragments of the comments~\cite{Stol2016}. We did not use a pre-defined coding schema, but interpreted comments in the context of the respective questions. 
Coding was first done individually and then consolidated in a sub-team. Another researcher crosschecked the consolidated coding and where necessary, the coding was adjusted in consensus. When reporting codes, Section~\ref{sec:results}, we provide a few examples using verbatim excerpts, including spelling and punctuation mistakes.

\section{Results}
\label{sec:results}

We present now the results of the data analysis. We start with demographics and then we zoom on the results per research question. 

\subsection{Experience and Demographics}
\label{subsec:demographic_information}
We received 113 valid responses (i.e., a response rate of 56\%). Based on the answer to the first question (Q0.1), we split the answers to the other questions of the demographics in two groups: those provided by all respondents and those provided by respondents that worked with concrete self-adaptive systems.  
\subsubsection{Experience with self-adaptation (Q0.1)}

Sixty two of the 113 participants (56\%) have worked with self-adaptive systems. This indicates that self-adaptation is frequently applied in practice. 

\subsubsection{Software systems built by organisations (Q0.2)}

The most frequent type of systems participants build in their company is \textit{Web/mobile systems} with 22 occurrences (19.5\%), 11 of them apply self-adaptation. Second comes \textit{embedded}, \textit{cyber-physical}, \textit{Internet of Things systems} (in particular \textit{robotic} and \textit{manufacturing}) with 16 occurrences (14.2\%), 10 of them applying self-adaptation. Both \textit{transportation} and \textit{networks} have 8 occurrences with 5 applying self-adaptation. The remaining systems include \textit{data management}, \textit{e-commerce}, and \textit{retail}. The results show that the participants work with a variety of systems and apply self-adaptation. This underpins the representatives of the data collected during the survey.  

\subsubsection{Software engineers working at companies (Q0.3)}
Fifty six of the 113 participants (49.5\%) work at a company with +100 software engineers; 30 of them apply self-adaptation. Thirteen participants (11.5\%) work at a company with less than 10 software engineers; seven of them apply self-adaptation. The remaining 44 participants (39\%) work in a company with 10 to 100 software engineers; 25 of them apply self-adaptation. The application of self-adaptation is similar for all categories. These numbers show that companies with a wide range of employed software engineers are represented.

\subsubsection{Roles of participants in their organisation (Q0.4)}
Around three in four participants indicated they have a single role in their organisation. The other participants indicated that they have two or more roles. The most frequent roles are project manager/coordinator with 49 occurrences (43.5\%), 31 of them apply self-adaptation, and programmer with 48 occurrences (42.5\%), 26 apply self-adaptation. Other roles are designer/architect (29, 13 applying self-adaptation), tester and maintainer (each 10, 5 of them applying self-adaptation), and operator (7, 4 applying self-adaptation). 
A variety of software engineering roles are represented in our sample and those that work with self-adaptive systems are more or less equally distributed.

\subsubsection{Experience of participants (Q0.5)}

Seventy six of the 113 participants (67.3\%) have more than nine years of experience as software engineer; 45 of them apply self-adaptation. Fifteen participants (13.2\%) have between one and three years of experience; five of them apply self-adaptation. Of the remaining 22 participants (19.5\%) with four to eight years of experience, 12 apply self-adaptation. These numbers show that most participants involved in the survey are mature software engineers and for all categories the representation of those that apply self-adaptation is similar. 

\subsection{Motivations for Self-adaptation (RQ1)}
\label{subsec:rq1}
We analyse now the data that we collected for answering RQ1. This research question focuses on the types of problems they solve using self-adaptation and what the benefits of self-adaptation could be. Note that the data we used to answer RQ1 and the other research questions comes from the 62 participants that have experience with self-adaptive systems (i.e., participants that answered ``Yes'' to Q0.1).

\subsubsection{For which problems do you apply self-adaptation? (Q1.1)}

On average, the participants applied self-adaptation for 3.7 problems (from a list with seven options), see Table~\ref{tab:q1-1}. Overall, practitioners apply self-adaptation to deal with a variety of problems. Optimising performance and automating tasks are the main problems tackled by self-adaptation (selected by 79\% of the participants). Other problems selected by at least half of the participants are (re-)configuring systems, and deal with changes in the environment. 

\begin{table}[b!]
\centering
\caption{Problems to apply self-adaptation (Q1.1). Percentages are fractions of participants that selected the problems}
\label{tab:q1-1}
\begin{tabular}{ll}
\hline\noalign{\smallskip}
Problem & Quantitative \\
\noalign{\smallskip}\hline\noalign{\smallskip}
To optimise performance & 49 (79\%)\\
To automate tasks & 41 (66\%)\\
To configure/reconfigure a system & 39 (63\%)\\
To deal with changes in the environment & 35 (56\%)\\
To detect and resolve errors & 30 (48\%)\\
To detect and protect a system against threats & 25 (40\%)\\
To deal with changes in the business goals & 8 (13\%)\\
Others & 10 (16\%)\\
\noalign{\smallskip}\hline
\end{tabular}
\end{table}

\begin{table*}[h!]
\caption{Analysis of answers - Reported benefits of self-adaptation (Q1.3)}
\label{tab:codes_q1-3}
\begin{tabular}{p{3.6cm}lp{13cm}}
\hline\noalign{\smallskip}
Categories/codes & \# & Example quotes\\
\noalign{\smallskip}\hline\noalign{\smallskip}
\rowcolor{lightgray} Improved utility & 48 &  \\
Robustness & 14 & ``fault tolerance, one node dies, a new one is spawned without manual intervention''; ``better error handling and prompt disaster recovery'' \\
Performance & 12 & ``Improve performance and quality-of-service''; ``increase in the speed of adaptation''  \\
Availability & 8 & 
``[...] 99.9999\% availability, which is crucial for some customers of these cloud-specific solutions''\\
Other qualities & 14 & ``for IoT: optimized operations, improved energy usage''; ``It is also an important part to guarantee the safety [...] of the overall system.''\\
\rowcolor{lightgray} Savings & 31 & \\
Costs & 17 & ``The primary benefit is cost reduction''; ``the cheaper bills for running this in an efficient manner [...]\\ 
Resources & 14 & ``scales down resources during hours when traffic is low, and scales up during peak hours, without any manual interference.''\\
\rowcolor{lightgray} Improved human interaction & 20  & \\
User experience & 12 & ``Keep Telco network in optimal condition so that QoS and user experience is maximized, and churn minimized''; ``better user satisfaction because of prompt website responses''\\
Burden engineers & 8 & ``removes most of the optimization burden from programmers, so they can be more productive''; ``Reduce workload on human operators; make (the results of) certain actions [...] repeatable and predictable''\\
\rowcolor{lightgray} Handle dynamics & 13 & \\
Context dynamics & 7 & ``Each machine is unique and its optimal operational parameters change over time due to ware, location, task and seasonal factor.''\\
Load dynamics & 6 & ``Change AGV behavior depending of the workload with the goal to save energy (battery life).''\\
\rowcolor{lightgray} Other improvements & 8  & \\
Various & 8 & `` In case of spikes in incoming events the system is able to adapt [...] avoiding bottlenecks.''; ``Easier and faster market integration''; 
\\
\noalign{\smallskip}\hline
\end{tabular}
\end{table*}

\subsubsection{What could be benefits of applying self-adaptation? (Q1.3)}

Fifty-eight respondents (94\% of those that worked with self-adaptive systems) provided useful descriptions of benefits of applying self-adaptation. On average, we identified 2.1 benefits per respondent. Table~\ref{tab:codes_q1-3} summarises the main findings. For each category (gray-shaded rows) and for each code per category we provide the number of occurrences and we give a few illustrative quotes of respondents. The dominating benefits of applying self-adaptation are improved utility (48 participants, i.e., 83\% of those that worked with self-adaptive systems) , in particular for robustness and performance, and savings in costs and resources (31 participants, 53\%).  

\begin{framed}
\noindent \textbf{Key insights from RQ1:} \vspace{3pt}\\
\noindent (1) Self-adaptation is primarily used in industry to optimise performance, automate tasks, configure/re-configure systems, and deal with changes in the environment. \\
\noindent (2) The main reported benefits of applying self-adaptation are improved utility and savings (in costs and resources), improved human interaction, and handling dynamics. \\
\noindent (3)  The problems tackled by industry are similar to those studied by academics. Yet,  practitioners do not put an emphasis on uncertainties, which is a key focus in research.\\ 
\noindent (4) The four classic tasks of self-adaptation studied by researchers (self-healing, self-optimising, self-protecting, and self-configuring) are also relevant to practitioners. 
\end{framed}

\subsection{Use Cases of Self-adaptation (RQ2)}
\label{subsec:rq2}

\subsubsection{Explain a concrete self-adaptive system you worked with (Q2.1)}

Sixty one participants (98\%) described a concrete self-adaptive system they worked with. 
Table~\ref{tab:codes_q2-2} summarises the findings. We identified three categories that characterise self-adaptive systems. Sixty one participants (98\%) described the subject of adaptation in their systems. System occurred 16 times, followed by module (i.e., a part of a system), and support system (infrastructure, platform, etc.), each with 12 times mentioned (20\%). Fifty two participants provided input to the types of adaptation they apply (81\%). Auto-scaling with 20 occurrences and automated reconfiguration with 10 occurrences are the most frequent types of adaptations applied by the participants. Finally, the participants explained in total 32 triggers for adaptation (52\%). The main triggers are changes in the (work) load of the systems and events, each with nine occurrences.  

\begin{framed}
\noindent \textbf{Key insights from RQ2:} 
\vspace{3pt}\\
\noindent (1) Self-adaptation is applied at different levels of software systems: from complete systems to parts and support systems. \\
\noindent (2) The dominating types of adaptation applied in industry are auto-scaling, automated re-configurations, and auto-tuning.  \\
\noindent (3)  Adaptions in industrial  systems are primarily triggered by dynamics in (work) load, the occurrence of events, and changes in properties of systems and their environments. \\ 
\noindent (4) Technologies such as elastic cloud,  auto-scaling, and container-orchestration systems such as Kubernetes, are key enablers for the realisation of self-adaptation in practice.  
\end{framed}

\begin{table*}[hbt]
	\caption{Analysis of answers -- Explain a concrete self-adaptive system you worked with (Q2.1)}
	\label{tab:codes_q2-2}
	\begin{tabular}{p{3.2cm}lp{13.2cm}}
		\hline\noalign{\smallskip}
		Categories and codes & \# & Example quotes\\
		\noalign{\smallskip}\hline\noalign{\smallskip}
		\rowcolor{lightgray} Subject of adaptation & 61 &  \\
		
		System & 16 & ``Autotuning of machine producing electricity.'' \\
		Module & 12 & ``Monitoring the memory/CPU/disk consumption of our servers and suggesting measures to fix it through human intervention.'' \\
		Support system & 12 & ``The HotSpot JVM of OpenJDK. It reads a program's Java bytecode, and adaptively tunes the performance of the program at runtime, adapting to runtime profiles.''; ``The adaptive integration platform (AIP)'' \\
		Cluster & 7 & ``auto-scaling of application cluster (computation resources) based on workload submitted by user - done in multiple environments'' \\
		Cloud & 7 & ``EBS (cloud auto-scaling). Monitors runtime performance of cloud instances and allows the establishment of automatic rules for scaling in/out the instances, e.g. responding to workload changes'' \\
		Application & 4 & ``auto-tuning systems (vehicles with driving support) [...] It could also adapt based on external factors like other vehicles.''\\
		Generic & 3 & ``We are currently developing a generic self-adaptation platform that can be applied to numerous cases. ''\\
		
		\rowcolor{lightgray} Type of adaptation & 52 &  \\
		
		Auto-scaling & 20 & ``Automated horizontal scaling of AWS EC2 instances for medical data processing systems''; ``autoscale a cluster based on the resource usage of the nodes of the cluster.''\\
		Automated reconfiguration & 10 & ``Our company develops safety critical systems for railway. Systems architecture is often with redundancy - e.g. 2 out of 3 system, where is automatic reconfiguration implemented. Purpose is high safety and availability.''\\
		Auto-tuning & 8 & ``A mink feeding robot, that can adjust the food amount according to a set of feeding rules and the food left over from last feeding. '' \\
		Monitor/Analysis & 6 & ``We configured AWS alarms to monitor performance of our systems in case we get more than few number of HTTP 400/500 errors''; ``Monitoring the memory/CPU/disk consumption of our servers and suggesting measures to fix it through human intervention.'' \\
		Automated CI & 3 & ``Continuos integration system - Monitors codebase \& starts building \& testing a new version as soon as it detects code changes 
		Build alignment - Creates a new release whenever a subsystem builds successfully'' \\
		Other & 5 & ``Proprietary load balancers that sits in front of bare metal servers.''\\
		
		\rowcolor{lightgray} Trigger of adaptation & 32 &  \\
		
		Load & 9 & ``Kubernetes, for handling load intensive periods for scaling up, and self recover from crashes.''; ``Autoscaling of SaaS applications in function of load on AWS and Azure clouds.'' \\
		Events & 9 & ``We use kubernetes which provides notification callbacks on any event such as host/pod not available, based on these events we auto mark the node was inactive and do not use those nodes for further write or read operations''; ``Auto Scaling an EMR cluster in AWS based on incoming event data''\\
		Environment properties & 6 & ``An IoT system running in Kubernetes and used to monitor water leaking for household insurance.''\\
		System properties & 6 & ``Auto-scaling functionality of an Azure Service Fabric cluster running a transformation load for processing AGV statistical and playback data.''\\
		User actions & 2 & ``[adapt] cache warm up strategy based on user interactions'' \\
		\noalign{\smallskip}\hline
	\end{tabular}
\end{table*}

\subsection{Difficulties with Self-Adaptation (RQ3)}
\label{subsec:rq3}

\subsubsection{Frequency of difficulties or challenges encountered when engineering or maintaining self-adaptive systems (Q3.1)} 

Twenty three respondents report that they sometimes face difficulties or challenges with applying self-adaptation (37\%). Nineteen frequently or very frequently encounter issues, while 7 rarely or very rarely have difficulties. One respondent reported to have always problems, while 6 reported that they have never problems. 

\subsubsection{Types of difficulties or challenges encountered when engineering or maintaining self-adaptive systems (Q3.2)}

Forty five respondents (73\%) reported on average 1.53 difficulties or challenges with engineering self-adaptive systems. Table~\ref{tab:codes_q4-2} summarises the findings. Most frequently reported issues (22 in total) relate to life cycle issues, in particular tuning/debugging. Design challenges were reported by 20 respondents, particularly reliable/optimal design. Other issues concern people and process issues and runtime challenges. 

\begin{framed}
\noindent \textbf{Key insights from RQ3:} 
\vspace{3pt}\\
\noindent (1) Seventy percent of the participants report that they face at least sometimes difficulties or challenges with self-adaptation. \\
\noindent (2) The difficulties and challenges crosscut the whole life cycle, from design and testing time to runtime and evolution. \\
\noindent (3) Top reported issues are tuning/debugging, reliable/optimal design and design complexity.  \\
\noindent (4)  Issues with skills and experience of people as well as process management are frequently reported. 
\end{framed}

\begin{table*}[hbt]
\caption{Analysis of answers - Difficulties and challenges encountered with engineering self-adaptive systems (Q3.2)}
\label{tab:codes_q4-2}
\begin{tabular}{p{3.8cm}lp{13.2cm}}
\hline\noalign{\smallskip}
Categories and codes & \# & Example quotes\\
\noalign{\smallskip}\hline\noalign{\smallskip}

\rowcolor{lightgray} Lifecycle issues & 22 &  \\

Tuning/debugging & 10 & ``Debugging the root cause of a scaling failure might be time-consuming: also, in some cases the problem might be outside of your control (e.g. temporary lack of EC2 Spot capacity in AWS)'' \\

Limitations of tools/methods & 7 & ``The metrics available are not always fully transparent and built with auto-scaling in mind''; ``IAM permissions are hard to deal with when configuring these self-adaptive systems. Usually, the permission to scale or to notify is not properly configured.''\\

System/environment evolution & 5 & ``If the functionality is not designed in from the beginning then it is a huge amount of work to implement later.''; ``System architecture over lifetime (nee features to be added...)'' \\

\rowcolor{lightgray} Design challenges & 20 &  \\

Reliable/optimal design & 10 & ``If things starts to fail - then we have major delivery problems very fast. This area is critical.''; ``the main challenge is to design adaptation function with respect to computation context - e.g., selecting right trade-off between shutting down instance or keeping it running longer time since boot of instance can be time-consuming.'' \\

Design complexity & 9 & ``Complexity in defining the adaptation rules. Conditions are not always obvious.''; ``Self-adaptiveness or resilience have to be taken into consideration at each stage of the software production and operation workflow of a distributed system. This is really a challenge as more often than not these are concepts that are completely obscure to the average programmer/devop mind.'' \\

Security & 1 & ``Software updates/dependencies, security'' \\

\rowcolor{lightgray} People and process issues & 14 &  \\

Skills/experience & 8 & ``Every self-adapt system must be tuned up which is sometimes tricky and needs high skilled engineers.''; ``The Kubernetes/Openshift cloud and centralized log storage (ElasticSearch - Logstash - Kibana) require experienced administration staff and vast knowledge of many networking concepts ([...], DNS, NAT'' \\

Process and management & 5 & ``We are not yet very experienced with the system, the main challenges were to convince the central IT department this was the way to go, then to design the system, and obviously master the technology'' \\

Documentation & 1 & ``humans don't write things down and don't like to revisit what they do to facilitate writing things down'' \\

\rowcolor{lightgray} Runtime challenges & 13 &  \\

Runtime uncertainty & 6 & ``Many self-adaptive systems are based on unproven heuristics. Therefore, they usually do not work in many cases.''; ``It is hard to guess how much can the environment affect the system. Usually, we develop the system on one or a limited set of units. It is hard to extend the parameters to cover whole production.'' \\

Data collection/evaluation & 5 & ``Gathering quantitative data samples to evaluate the performance is very complicated.''; ``sensors gives wrong reading values'' \\

Delayed/missing changes & 2 & ``Autoscaling is often too slow or triggered too late.'' \\

\noalign{\smallskip}\hline
\end{tabular}
\end{table*}

\section{Threats to Validity}~\label{sec:discussion}\vspace{-5pt}

The sample of practitioners that participated in this survey were recruited from our networks with industry. Since several of the researchers involved in this study are active in the field of engineering self-adaptive systems, the  practitioners of these networks may have been biased and inclined to apply self-adaptation more often. To anticipate this validity threat, we did not particularly focused on practitioners we have worked with in projects, but invited practitioners in various software engineering roles that are active across a wide range of domains. 

Further, our sample is not well balanced on a global scale, hence it may not be possible to generalise the results to industry in parts of the world that are only marginally represented in the sample. To anticipate this threat, we are currently running a second round of our survey to compensate for underrepresented areas to guarantee a better geographic coverage. 

We used the term self-adaptation to formulate questions about systems (or parts) that are equipped with a feedback loop. Hence, most questions required basic knowledge of the concept of self-adaptation. Respondents may have misinterpreted the concept or some of the questions. To anticipate this threat, we have carefully introduced the notion of self-adaptation in the survey, leveraging on the input we received from validating the questionnaire. 

Since our sample for the survey was drawn from practitioners of our networks with industry, they may not be representative for the population of software engineers in general and those that apply self-adaptation in particular. However, the results of the demographics of our sample show that the respondents were active practitioners with sufficient expertise in various roles across companies of different sizes.  

Regarding the qualitative analysis of answers with free text we used coding. Coding is a creative task that is influenced by the experience (and even opinions) of the coders~\cite{Fernandez2016}. To mitigate potential subjectivity, the data analysis was performed by multiple researchers and the codes and concepts that emerged were discussed to reach consensus.

\section{Conclusions and Future Work}~\label{sec:conclusions}\vspace{-5pt}

This paper reports preliminary results of a survey that aims at understanding the state of the practice in the use of self-adaptation. We focused on the motivations of practitioners to apply self-adaptation, concrete use cases, and  difficulties practitioners face when applying self-adaptation. The results show that practitioners apply self-adaptation for problems similar to those that are studied in academics. Yet, practitioners put more emphasis on costs and resources (compared to an emphasis on quality properties by researchers) and less on uncertainty (which is a key concern in academics, see e.g.,~\cite{Ramirez2012,Esfahani2013,Mahdavi2017,Hezavehi21}). Self-adaptation is applied at different levels of granularity of the system, and particular emphasis is given to system support (deployment and execution environment, integration platforms, etc.). The main focus is on auto-scaling, automated reconfiguration, and auto-tuning, exploiting key technologies such as elastic cloud and container-orchestration. 
A majority of practitioners face both technical difficulties and challenges with realising self-adaptation as people and process issues. 

The results reported in this paper are part of a large-scale study that is ongoing at the time of writing. We recently launched a second round of the survey with 100 additional practitioners. Besides the topics reported in this paper, we collect also data about \textit{how} practitioners realise self-adaptation, how they ensure \textit{trustworthiness} of their solutions, what \textit{risks} they face and how they mitigate these risks, what future \textit{opportunities} they see for self-adaptation, and how \textit{collaboration} with academics could help them with using self-adaptation in practice. In addition, we are running a series of interviews with practitioners that reported not to apply self-adaptation to better understand the reasons for this.  

\balance



\section*{Acknowledgment}

We thank the participants that contributed input to the study, the reviewers of the survey protocol, and Linnaeus University for hosting the survey. The financial support by Christian Doppler Research Association, the BMDW and the National Foundation for Research, Technology and Development is gratefully acknowledged.





\bibliographystyle{ACM-Reference-Format}
\bibliography{references}



%







\end{document}